\documentclass{aa}

\usepackage{graphicx}
\usepackage[varg]{txfonts}

\usepackage{natbib}
\bibpunct{(}{)}{;}{a}{}{,}

\usepackage{amssymb}

\usepackage{textcomp}

\usepackage[mathlines,switch]{lineno}

\newcommand{\gray}{\mbox{$\gamma$-ray}\xspace}
\newcommand{\threec}{\mbox{3C\,111}\xspace}

\newcommand{\fgamma}{F-GAMMA\xspace}

\begin{document}

\title{Sub-milliarcsecond imaging of a bright flare and ejection event in the extragalactic jet 3C\,111}

\author{Robert~Schulz\inst{\ref{inst2}, \ref{inst3}, \ref{inst1}}
	\and M.~Kadler\inst{\ref{inst2}}
	\and E.~Ros\inst{\ref{inst4}}
	\and M.~Perucho\inst{\ref{inst5},\ref{inst6}}
	\and T.~P.~Krichbaum\inst{\ref{inst4}}
	\and I.~Agudo\inst{\ref{inst7}}
	\and T. Beuchert\inst{\ref{inst8},\ref{inst9},\ref{inst3},\ref{inst2}}
	\and M. Lindqvist\inst{\ref{inst10}}
	\and K.~Mannheim\inst{\ref{inst2}}
	\and J. Wilms\inst{\ref{inst3}}
	\and J. A. Zensus\inst{\ref{inst4}}
}

	\institute{Lehrstuhl f\"{u}r Astronomie, Universit\"{a}t W\"{u}rzburg, Campus Hubland Nord, Emil-Fischer-Strasse 31, 97074 W\"{u}rzburg, Germany, \email{r.f.schulz@issc.leidenuniv.nl, matthias.kadler@astro.uni-wuerzburg.de}\label{inst2}
	\and Dr. Remeis Sternwarte \& ECAP, Universit\"{a}t Erlangen-N\"{u}rnberg, Sternwartstr. 7, 96049 Bamberg, Germany\label{inst3}
	\and Netherlands Institute for Radio Astronomy (ASTRON), Oude Hoogeveensedijk 4, 7991 PD Dwingeloo, The Netherlands\label{inst1}
	\and Max-Planck-Institut f\"{u}r Radioastronomie, Auf dem H\"{u}gel 69, 53121 Bonn, Germany\label{inst4}
	\and Departament d'Astronomia i Astrof\'{\i}sica, Universitat de Val\`{e}ncia, 46100 Burjassot, Val\`{e}ncia, Spain\label{inst5}
	\and Observatori Astron\`omic, Universitat de Val\`encia, Parc Cient\'{\i}fic, C. Catedr\'atico Jos\'e Beltr\'an 2, 46980 Paterna, Val\`encia, Spain\label{inst6}
	\and Instituto de Astrof\'{i}sica de Andaluc\'{i}a, CSIC, Apartado 3004, 18080, Granada, Spain\label{inst7}
	\and European Southern Observatory, Karl-Schwarzschild-Str. 2, 85748 Garching bei München, Germany\label{inst8}
	\and Anton Pannekoek Institute for Astronomy, P.O. Box 94249, 1090GE Amsterdam, The Netherlands\label{inst9}
	\and Department of Earth and Space Sciences, Chalmers University of Technology, SE-43992 Onsala, Sweden\label{inst10}
}

\date{Received date / Accepted date}

\abstract
{Flares in radio-loud active galactic nuclei (AGN) are thought to be associated with the injection of fresh plasma into the compact jet base.
 Such flares are usually strongest and appear earlier at shorter radio wavelengths. Hence, very long baseline interferometry (VLBI) at mm-wavelengths is best suited to study the earliest structural changes of compact jets associated with emission flares.}
{We study the morphological changes of the parsec-scale jet in the nearby ($z=0.049$) gamma-ray bright radio galaxy \threec following a flare that developed into a major radio outburst in 2007.}
{We analyse three successive observations of \threec at 86\,GHz with the Global mm-VLBI Array (GMVA) between 2007 and 2008 which yield a very high angular resolution of $\sim45\mathrm{\,\mu as}$. In addition, we make use of single-dish radio flux density measurements from the F-GAMMA and POLAMI programmes, archival single-dish and VLBI data.}
{We resolve the flare into multiple plasma components with a distinct morphology resembling a bend in an otherwise remarkably straight jet. The flare-associated features move with apparent velocities of $\sim4.0$\,c to $\sim4.5$\,c and can be traced also at lower frequencies in later epochs. Near the base of the jet, we find two bright features with high brightness temperatures up to $\sim 10^{11}\mathrm{\,K}$, which we associate with the core and a stationary feature in the jet.}
{The flare led to multiple new jet components indicative of a dynamic modulation during the ejection. We interpret the bend-like feature as a direct result of the outburst which makes it possible to trace the transverse structure of the jet. In this scenario, the components follow different paths in the jet stream consistent with expectations for a spine-sheath structure, which is not seen during intermediate levels of activity. The possibility of coordinated multiwavelength observations during a future bright radio flare in \threec makes this source an excellent target for probing the radio-\gray connection.}


\maketitle

\section{Introduction}
\label{Intro}
	
	A common signature of relativistic jets in radio-loud active galactic nuclei (AGN) is their strong variability in radio emission (e.g., \citealt{Laehteenmaeki1999,Hovatta2008,Hovatta2009,Richards2014}). This is generally explained by shocks traveling through the parsec-scale radio jet leading to the appearance of new features in high-resolution images on milliarcsecond (mas) scales (e.g., \citealt{Marscher1985,Valtaoja1999,Savolainen2002,Fromm2013a}). Moreover, combined studies of radio and \gray{} light curves reveal a correlation of flares occurring in both energy regimes with the radio emission often seen first at mm-wavelengths (e.g., \citealt{Laehteenmaeki2003,Leon-Tavares2011a,Fuhrmann2014}).
	In the TeV \gray{} band, extreme variability on time scales as short as minutes has been observed (\citealt{Albert2007,Aharonian2007,Aleksic2014a}), which suggests extremely small and compact substructure in AGN jets.
	Rapid broadband variability is not exclusive to blazars, but can also be seen in radio galaxies with small viewing angles of the jet to the line of sight (e.g., \citealt{Jorstad2001b,Marscher2010,Agudo2011,Schinzel2012,Karamanavis2015b,Casadio2015b}). In fact, the favorable orientation and typically small distances allow us to probe substantially smaller linear scales in high-resolution imaging studies of radio-galaxy jets than in blazars (e.g., \citealt{Boccardi2017}).
	
	An important open question concerns the physical origins of high-energy blazar variability. These are likely related to changes of particle acceleration and cooling processes in the most compact regions of the jets. Various propagation effects likely play a role but
	cannot be probed by any direct means due to the limited angular resolution at \gray{} energies.
	Frequent VLBI observations of \gray{}-bright radio galaxies at mm-wavelengths as made possible with the Global mm-VLBI Array (GMVA\footnote{\url{http://www.mpifr-bonn.mpg.de/div/vlbi/globalmm/}}) in combination with single-dish flux density monitoring are best suited to perform such a study as they provide angular resolutions down to $\sim40\mathrm{\,\mu as}$, which can correspond to sub-parsec linear deprojected scales for nearby sources.
	
	One prime target for such studies is the nearby radio galaxy \threec{} at a redshift $z=0.049$ \citep{Veron-Cetty2010}. The kpc-scale radio morphology \citep{Linfield1984} is consistent with the Fanaroff-Riley class II \citep{Fanaroff1974}. 
	The parsec-scale jet exhibits apparent superluminal motion up to  speeds of $\sim8c$ and the angle of the jet to the line of sight has been estimated between $\sim10\degr$ and $\sim20\degr$ \citep[although smaller angles have not been entirely excluded;][]{Goetz1987,Jorstad2005,Lewis2005,Kadler2008,Hogan2011,Lister2013,Beuchert2018}. \threec{} has shown several radio flares in the past with two exceptional long-lasting outbursts of several months above 10\,Jy in flux density in the 3\,mm-band \citep{Alef1998,Trippe2011,Chatterjee2011}. The first one occurred in early 1996 \citep{Alef1998} and the second one in the middle of 2007 \citep{Chatterjee2011}. Both developed into complex features in the jet at 15\,GHz and 43\,GHz that differed from other components related to lower-activity phases (\citealt{Kharb2003,Jorstad2005,Kadler2008,Chatterjee2011,Beuchert2018,Jorstad2017}). \threec{} has also been detected at \gray{} energies and seems to be a rather faint \gray{} emitter most of the time and it becomes bright and detectable only during short flaring periods 
	\citep{Hartman2008,Grandi2012}. It has been concluded that the site of the \gray{} flares of \threec is confined to a compact region smaller than 0.1\,pc inside the unresolved cm-VLBI core at distances of not more than 0.3\,pc from the central black hole.
	
	This paper presents results of very high angular resolution observations with the GMVA over a period of one year during a major radio outburst.
	\cite{Schulz2013} presented preliminary images, but here we present the final images based on the fully calibrated data and subsequent results. The next section describes the radio data and their reduction (Sect. \ref{Data}). Section \ref{Results} and Sect. \ref{Discussion} present our results and their discussion, which is followed by a summary in Sect. \ref{Summary}.
	
	Throughout this paper, we adopt the $\Lambda$CDM cosmology with the parameters \mbox{$H_0=70\,\mathrm{km\,s}^{-1}\,\mathrm{Mpc^{-1}}$}, $\Omega_M=0.3$, and $\Lambda=0.7$ \citep{Freedman2001}. Hence, at the redshift of \threec{}  we have a projected linear scale of $1\mathrm{\,mas}\approx 0.96\mathrm{\,pc}$. Assuming an inclination of the jet of $13^\circ$ (see above), the angular beam of the GMVA of (40 -- 70)\,$\mu$as can spatially resolve linear deprojected scales down to $\sim 0.17$\,pc. The linear resolution corresponds to about 9900\,Schwarzschild radii assuming a mass of $1.8^{+0.5}_{-0.4}\times 10^8M_\sun$ for the central supermassive black as determined by \cite{Chatterjee2011}.

\section{Observation and Data Reduction}
\label{Data}

	\subsection{GMVA observations}
	\label{Data:GMVA}
	
		Following the outburst of \threec in 2007, we performed three successive observations with the GMVA at 86\,GHz on 2007 Oct 15, 2008 May 11 and 2008 Oct 14 (experiment code GK039A,B and GK040). The configuration of the GMVA at the time of our observations comprised five European and eight North-American telescopes, which are listed in Table \ref{tab:Data:GMVA}.
		
		The GMVA data were correlated with the VLBI correlator at the Max Planck Institute for Radio Astronomy (MPIfR, Germany). We used the \textsl{Astronomical Imaging Processing System} (\textsf{AIPS}, \citealt{Greisen2003}) to perform the a-priori amplitude and phase calibration. For this purpose, we corrected for atmospheric attenuation at each station which is a critical issue for mm-VLBI because of the increasing opacity of the atmosphere at high radio frequencies. In addition, measurements of the antenna's system temperature and gain factors were applied. The phases were calibrated using the global fringe fitting algorithm \citep{Schwab1983}.
		
		The 86\,GHz GMVA data were further self-calibrated and imaged with the software package \textsf{DIFMAP} \citep{Shepherd1997}. An initial model was created using the \textsf{CLEAN} algorithm \citep{Hoegbom1974} and phase self-calibration combined with flagging of bad data points which was followed by amplitude self-calibration over the whole observing time. This procedure was repeated with subsequent smaller time intervals for amplitude self-calibration. The images were produced by using the final \textsf{CLEAN}-model convolved with the \textsf{CLEAN}-beam, i.e., a two-dimensional Gaussian approximation of the dirty beam, and adding the residual noise. The hybrid imaging process was performed with natural weighting of the visibilities. The properties of the resulting images are listed in Table \ref{tab:Data:GMVA-Images}.
		
		We tested the absolute amplitude calibration by comparing the total flux density with quasi-simultaneous single-dish measurements from the \fgamma{} programme at 86\,GHz (2007 Oct 9: $12.06\pm0.22\mathrm{\,Jy}$, 2008 May 2: $4.05\pm0.16\mathrm{\,Jy}$, 2008 Oct 7: $4.06\pm0.21\mathrm{\,Jy}$). We consistently use F-GAMMA measurements (see Sect. 2.2) for our amplitude-calibration check and scaling, which benefit from simultaneous multifrequency measurements that reduce the risk of systematic gain offsets. We also tested the independent additional adjacent POLAMI measurements, which lead to consistent results within the F-GAMMA measurement uncertainties.
		
		We found a significant difference between the single-dish and VLBI flux density for the first two observations by a factor of $1.5\pm0.2$ and $2.1\pm0.2$, respectively. It is unlikely that this discrepancy stems from missing flux density resolved out by VLBI because of the high observing frequency. In addition, the last observation is consistent with the single-dish measurements ($1.1\pm0.2$). Hence, we consider uncertainties in the system temperature measurements and the gain curves of the telescopes to be the likely origin of the inconsistency between VLBI and single-dish flux density of the first two observations. We empirically corrected for this by scaling the \textsf{CLEAN}-model from the initial imaging by the factors given above. The new \textsf{CLEAN}-model was used to determine a constant amplitude-correction factor for each telescope. The scaled visibilities were then imaged and self-calibrated again. For the initial images, we estimated an uncertainty of the total flux density of about 15\% based on repeating the imaging process several times. This uncertainty applies to the scaled data and the resultant images. For the third GMVA observation, we estimated an additional systematic uncertainty of the flux density, because the 86\,GHz lightcurve suggests that this VLBI observation occurs during a rise in flux density. Based on an interpolation between the two adjacent F-GAMMA observations, we find that the flux density of the third observation might be underestimated by about 22\%. Because of uncertainty in the actual lightcurve evolution, we prefer to use the scaling factors from real flux density measurements. Therefore, we did not scale the third VLBI observation as discussed above, but we indicate the systematic uncertainty for the third observation where necessary in this paper.
		
		The self-calibrated data were modeled with two-dimensional circular Gaussian components using the fit routine implemented in \textsf{DIFMAP}. The aim of this process is to find a consistent description of the evolving brightness distribution of the jet throughout the three observations.
		The uncertainties of the parameters for each component were first calculated with a purely statistical approach based on \cite{Fomalont1999} which may underestimate the true uncertainties because this approach does not account for systematic errors introduced during the aperture-synthesis processing, i.e., by data editing, deconvolution, and (self-)calibration. Hence, we conservatively alter the uncertainties for the flux density of the components to the 15\,\% level adopted from amplitude calibration. Similarly, we alter the uncertainties for the positions of the components to 50\,\% of the component's major axis. The best-fit parameters of all components are listed in Table \ref{tab:Components}.

	\subsection{Ancillary data}
	\label{Data:Ancillary}
	
		The \fgamma{} (\textit{Fermi-GST} AGN Multi-frequency Monitoring Alliance) programme\footnote{\url{https://www3.mpifr-bonn.mpg.de/div/vlbi/fgamma/fgamma.html}} monitored a sample of \gray-bright blazars with the Effelsberg 100-m and the Pico Veleta 30-m radio telescopes over a broad range of radio frequencies from January 2007 till January 2015 \citep{Angelakis2010,Angelakis2012,Fuhrmann2014,Fuhrmann2016b,Angelakis2019}. Here, we focus on measurements at 86\,GHz obtained with the Pico Veleta telescope between 2007 and 2009 (see Table \ref{tab:lightcurve_data}).

		Additional flux density measurements of the source at 86\,GHz (see Table \ref{tab:lightcurve_data}), published in \cite{Agudo2014}, are provided by the POLAMI (Polarimetric Monitoring of AGN at Millimeter Wavelengths) programme\footnote{\url{http://polami.iaa.es}} (\citealt{Agudo2018a,Agudo2018b,Thum2018}). The data were also taken with the IRAM 30\,m telescope at the Pico Veleta Observatory in the same time frame as the F-GAMMA programme. The observing strategy and data reduction of the POLAMI programme is described in detail in \cite{Agudo2018a}.

		We further make use of selected VLBA data at 43\,GHz that are available in a fully calibrated state from the VLBA-BU-BLAZAR monitoring programme\footnote{\url{https://www.bu.edu/blazars/VLBAproject.html}} (e.g., \citealt{Marscher2011,Jorstad2017}). These data were previously discussed by \cite{Chatterjee2011} and \cite{Jorstad2017}. Similar to the GMVA data, we fit two-dimensional Gaussian components to the visibilities to represent the brightness distribution of the jet. We selected observations which were close in time to the GMVA data. We also make use of previously published SMA flux density measurements at 230\,GHz over a similar time range to the \fgamma{} data \citep{Chatterjee2011}.

\section{Results}
\label{Results}

	\begin{figure}[t!]
		\includegraphics[width=1\linewidth]{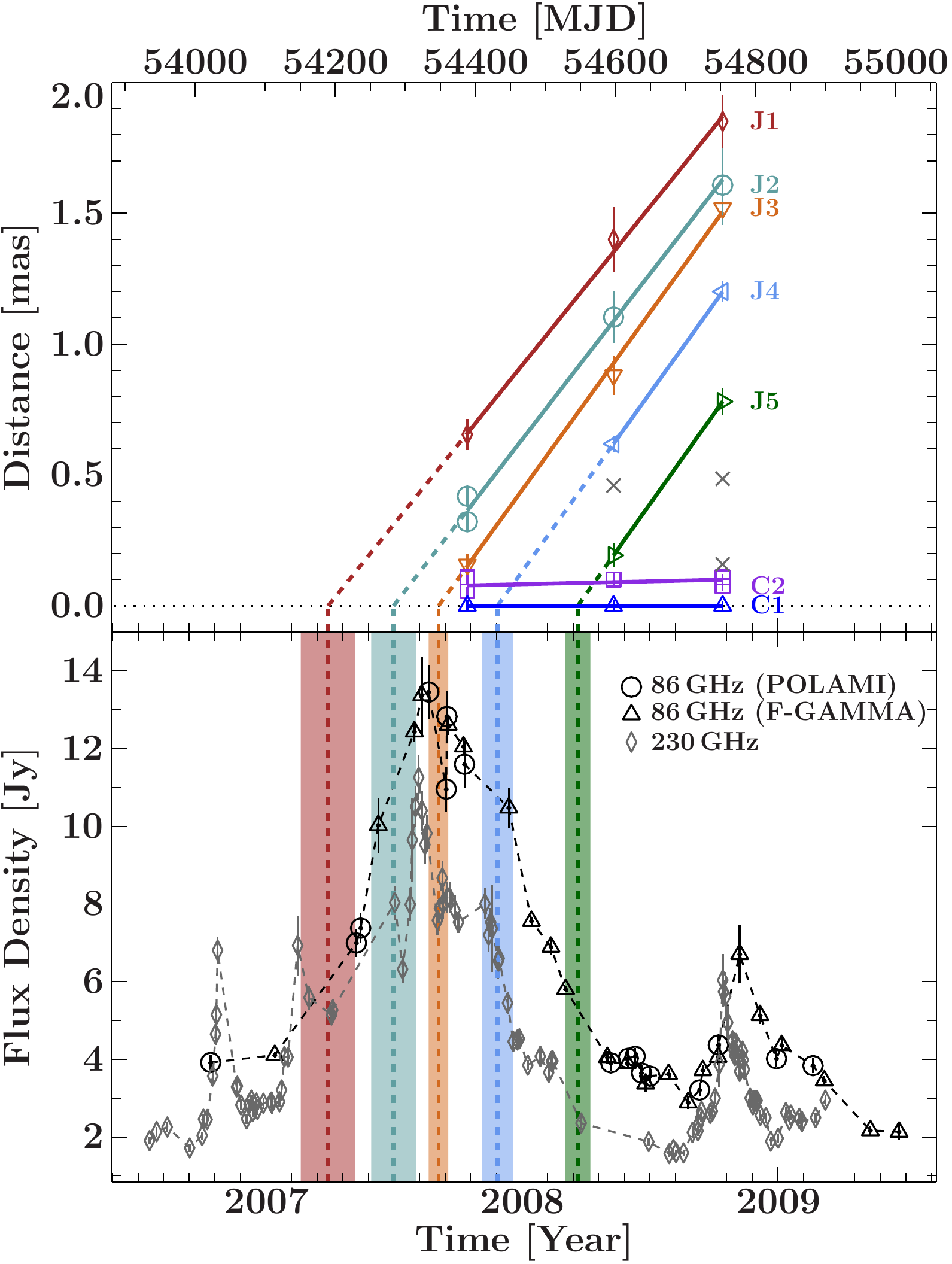}
		\caption{Upper panel: distance of jet features to the central region over time. Highlighted in color are the associated components and in gray unassociated modelfit components. The solid lines were calculated from the vector fit of the apparent transverse velocity. The colored dashed lines represent extrapolations to the estimated ejection times in the distance-time-domain. 
		Bottom panel: radio light curve at 86\,GHz (F-GAMMA, POLAMI) and 230\,GHz \citep{Chatterjee2011} between 2006.5 and 2009.5.
		The ejection times of the components are highlighted as vertical, colored, dashed lines in both panels. The wide colored bands indicate the uncertainty in the ejection time of the components.}
		\label{fig:86Ghz-DistTimeLC}
	\end{figure}

	\subsection{Single-Dish light curve}
	\label{Results:LC}
		
		The 86\,GHz and 230\,GHz light curves between 2006.5 and 2009.5 are shown in the bottom panel of Fig.~\ref{fig:86Ghz-DistTimeLC}. The peak of the major outburst occurred around August 2007 and was followed by a smaller flare in late 2008 before returning to a more typical lower state. 
		
		The first GMVA observation was conducted only two months after the peak in the light curve at which time \threec{} was clearly still in a very high state. 
		The second GMVA observation was performed in a state when the single-dish light curve had already dropped to about half the peak value. The third GMVA observation took place close to the peak of the secondary smaller flare in late 2008 and the single-dish flux density measurement reached the same value as around the second GMVA epoch.
	
	\subsection{Morphology and time evolution of the 86-GHz jet}
	\label{Results:Images}
	
\begin{figure}
			\includegraphics[width=.9\linewidth]{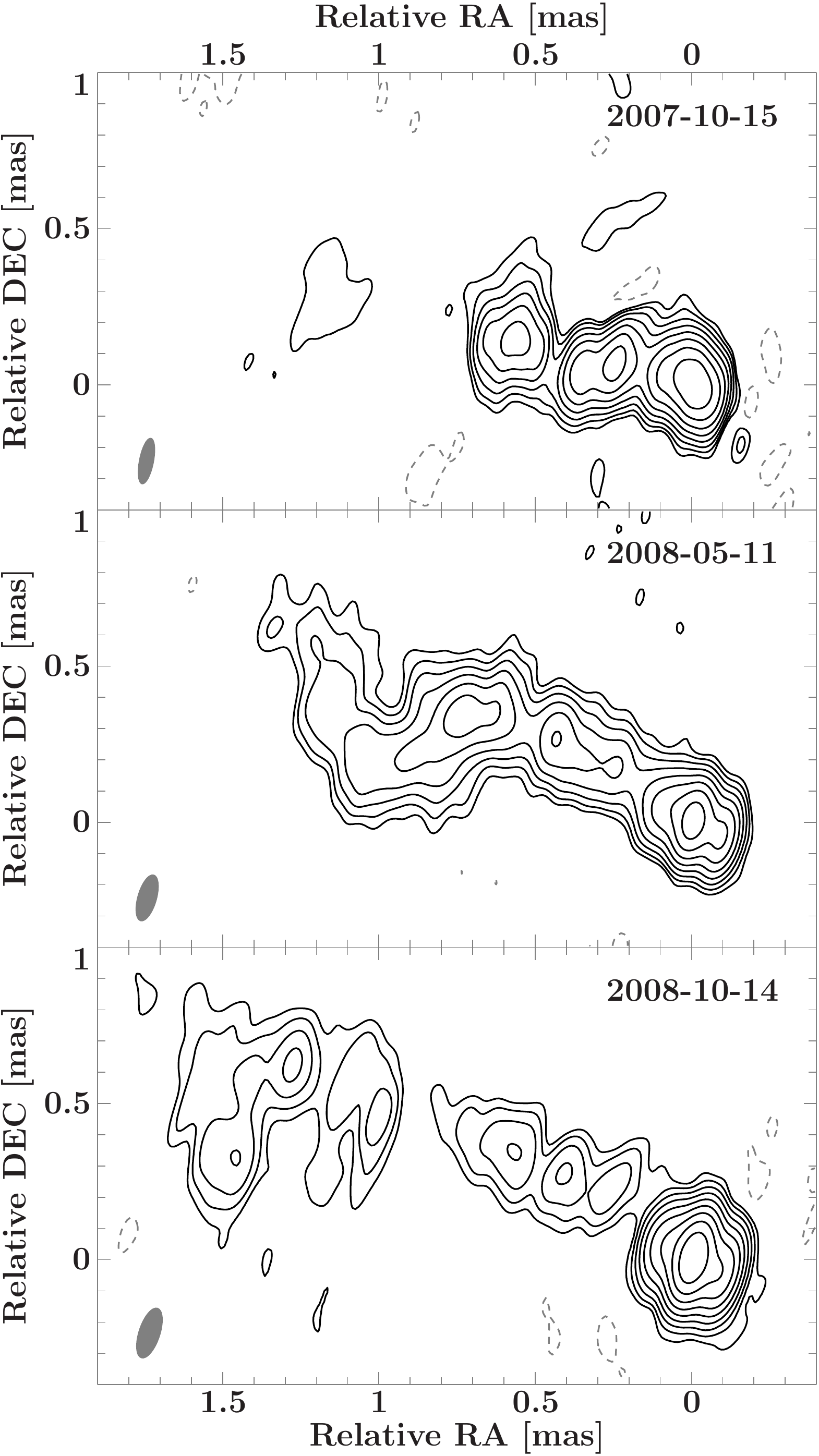}
			\caption{Images from the three GMVA observations of \threec (image parameters given in Table \ref{tab:Data:GMVA-Images}). Contour lines start at three times the individual $1\sigma$ noise level and increase logarithmically by factors of 2. Negative fluctuations at the $-3\sigma$-level are indicated by dashed, gray contour lines. The gray shaded ellipse in the lower left corner represents the synthesized beam.}\hspace{.1\linewidth}
			\label{fig:86Ghz-Images}
		\end{figure}

		Figure~\ref{fig:86Ghz-Images} shows the three images obtained by the GMVA observations. They are centered on the brightest feature and reveal a single-sided jet starting in the east-northeast direction. The first and second observations show an unusual change in the position angle of the jet at a distance of $\sim 0.5\mathrm{\,mas}$ and $\sim 1\mathrm{\,mas}$, respectively. The jet seems to form a bend which increases in scale between both observations. By the time of the third observation the bend has evolved into a larger and more diffuse complex emission region. We also find faint diffuse emission at a distance around 4\,mas, which lies outside of the chosen plot range of Fig. \ref{fig:86Ghz-Images} as it is not relevant for this study.
		
\begin{figure}
			\centering
			\hspace{.15\linewidth}\includegraphics[width=.85\linewidth]{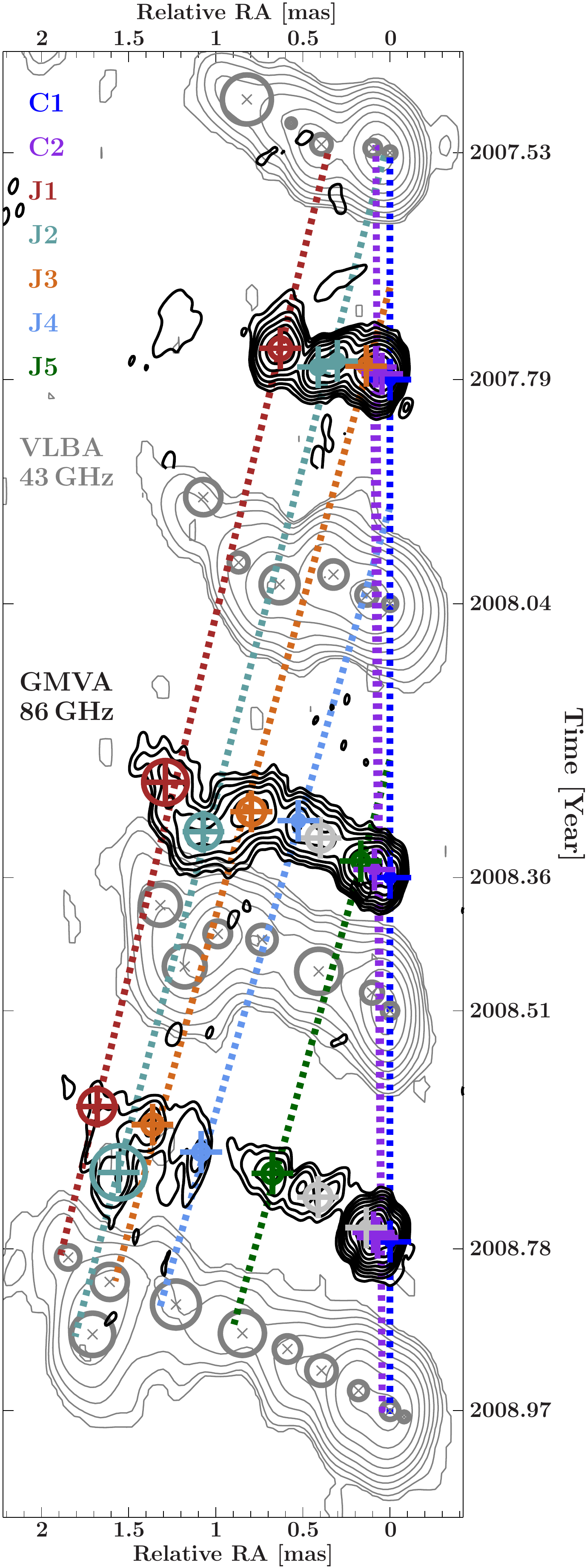}
			\caption{Time evolution plot of the source morphology at millimeter wavelengths. The black solid contour lines are based on the \textsf{CLEAN}-images at 86\,GHz (GMVA) after restoring with a common beam ($0.167\mathrm{\,mas}\times0.067\mathrm{\,mas}$, $-18.2\degr$) and contour lines starting at three times the noise level. Gaussian model components are superimposed on top of the \textsf{CLEAN}-images. The images and components are aligned to the position of C1. The gray dashed contour lines show 43\,GHz images from the Boston-University blazar group monitoring programme aligned to the position of the core based on model fitting. The colored, dashed lines were calculated from the kinematic fit to the GMVA components and extrapolated to the first and last 43\,GHz observation shown here.}
			\label{fig:86Ghz-TimeEvo}
		\end{figure}

		In order to better quantify the morphology of the jet and its changes over time, we use the fit parameters of the Gaussian modelfit components. The large time span of the GMVA observations, the fast evolution of the jet and the limited number of observations make it difficult to robustly cross-identify the Gaussian modelfit components fitted to the three observations. The cross identification of the components is mainly done using morphological similarities and through comparison with near-in-time VLBI images at 43\,GHz which exhibit similar jet features. The resulting time evolution is shown in Fig.~\ref{fig:86Ghz-TimeEvo}. 
				
		In all three 86-GHz observations, we find that the most western component which we label as `C1' does not correspond to the brightest feature of the jet. We consider it to be the stationary VLBI core and not a sign of a possible counter-jet (see discussion in Sect. \ref{Discussion:Core}).	Therefore, it is used to align the modelfit components at 86\,GHz in the subsequent analysis. We label the brightest feature of the jet as component `C2'. In the first and last GMVA observations, it is modeled with two, nearby Gaussian model components which are both necessary to represent the associated visibility data, thus we treat them as a single entity. For the remaining components of the jet, we assign a `J' plus an integer that increases with the (back-extrapolated) ejection time (see Fig.~\ref{fig:86Ghz-DistTimeLC}, top panel). In the first epoch, J2 consists of two Gaussian model components which we consider to trace the same emission region due to their proximity to each other.
		
\begin{figure}
			\includegraphics[width=1\linewidth]{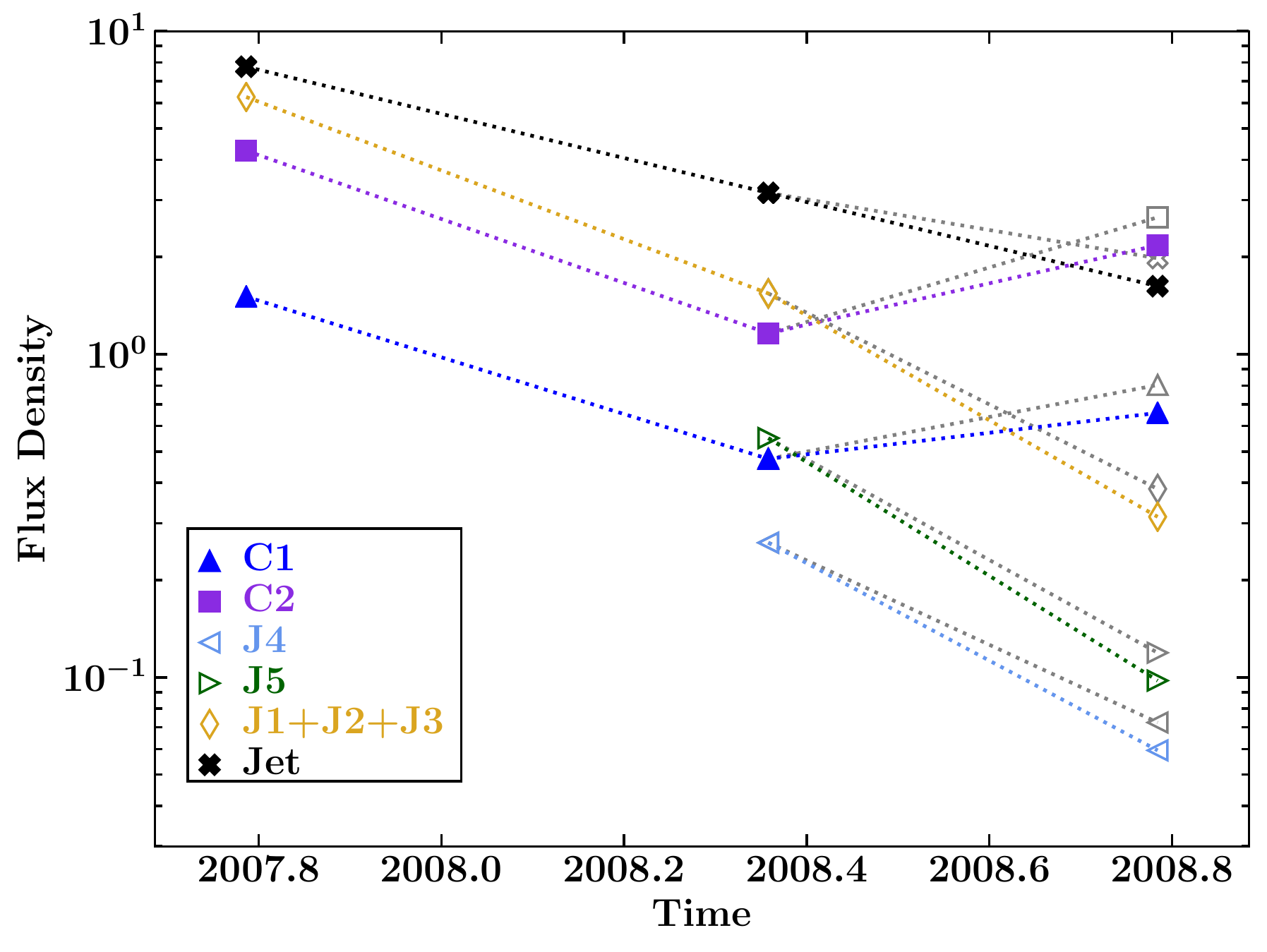}
			\caption{Flux-density light curves of \threec based on the GMVA data separated into C1, C2, J4, J5, and the sum of the components associated with the bend (J1, J2, J3) and the sum of all components representing the jet, i.e., all components except C1 and C2.
			For the third VLBI observation, indicated in gray the flux density of the components if these were scaled using the interpolated scaling factor (see Sect. \ref{Data:GMVA}) as a systematic uncertainty.}
			\label{fig:86Ghz-FluxEvoComp}
		\end{figure}

		Figure \ref{fig:86Ghz-FluxEvoComp} shows the resulting flux density evolution of C1, C2 and the jet components. There are significant changes in the flux-density distribution over time. In the first observation, $\sim 48\%$ of the total flux density is associated with C1 and C2. This fraction decreases slightly in the second observation to $\sim 37\%$ before increasing to $\sim 74\%$ in the third epoch.
		This behavior is consistent with the evolution of the single-dish light curve (Fig. \ref{fig:86Ghz-DistTimeLC}), which indicates that the third GMVA observation coincides with the onset of the secondary smaller flare in late 2008. This flare seems to be localized to C1 and C2. The brightness temperatures of C1 and C2 are of the order of $10^{11}\mathrm{\,K}$ in all three epochs (see Table~\ref{tab:Components}). 
	
		We fit the two-dimensional position of the components over time with linear regression in order to determine the angular velocities $v_x$ and $v_y$. Based on this, we calculated the apparent angular velocities $v_\mathrm{app}$. The ejection time was determined from the linear regression in the distance-time domain. The results are listed in Table \ref{tab:Results:Kinematic}. Care has to be taken in the interpretation of these estimates, considering the limited number of observations at 86\,GHz.
		
		The results suggest that C2 is a stationary component consistent with the interpretation of \cite{Jorstad2017} which has strong implications for the jet properties at 15\,GHz \citep{Beuchert2018}. We are able to trace the bend in the jet with two distinct components, J2 and J3, which move with an apparent velocity of $\sim4.2c$ and $\sim4.5c$, respectively. Similar values are also estimated for the other components. We estimate the critical angle $\theta_\mathrm{LOS,crit}$ of the jet to the line of sight (e.g., \citealt{Cohen2007}) that corresponds to the maximum of $\beta_\mathrm{app}$ with $\cos\theta_\mathrm{LOS,crit}$ = $\beta_\mathrm{app} / \sqrt{1+\beta_\mathrm{app}^2}$ to be $\theta_\mathrm{LOS,crit} \sim 13\degr$ consistent with previous estimates at lower frequencies. This leads to a critical Doppler factor $\delta_\mathrm{crit}=\gamma_\mathrm{min} \sim 4.5$.
		
		The position angle (PA) of J2 and J3 differs by $\sim 10\degr$ and J1 shows a change in PA of $\sim 10\degr$ over time (see Table \ref{tab:Components}). The median PA for all components excluding C1 and C2 is about 65\degr which is consistent with observations at 15\,GHz and 43\,GHz \citep{Beuchert2018,Jorstad2017}.
		
		\begin{table*}
		\centering
		\caption[]{Properties of the 86\, GHz images}
		\label{tab:Data:GMVA-Images}
		\begin{tabular}{ccp{3.5cm}cccccc}
		\hline
		Date\tablefootmark{a} & Date\tablefootmark{a} & GMVA configuration\tablefootmark{b} & $\sigma_\mathrm{noise}$\tablefootmark{b} & $S_\mathrm{peak}$\tablefootmark{c} & $S_\mathrm{total}$\tablefootmark{d} & $b_\mathrm{maj}$\tablefootmark{e} & $b_\mathrm{min}$\tablefootmark{e} & $b_\mathrm{PA}$\tablefootmark{e}\\
		$[$YYYY-MM-DD$]$ & [year] &  & [mJy\,beam$^{-1}$] & [Jy\,beam$^{-1}$] & [Jy] & [$\mu$as] & [$\mu$as] & [$\degr$] \\
		\hline
		\hline
		2007-10-15  & 2007.79 & Ef-On-Me-PdB-VLBA & $0.32$ & $2.41\pm0.36$ & $11.8\pm1.8$  & $148$ & $45$ & -11\\
		2008-05-11  & 2008.36 & Ef-On-PV-VLBA & $0.16$ & $0.93\pm0.14$ & $4.27\pm0.64$ & $152$ & $59$ & -16\\
		2008-10-14  & 2008.78 & Ef-On-VLBA & $0.13$ & $1.87\pm0.28$\tablefootmark{f} & $3.78\pm0.57$\tablefootmark{f} & $167$ & $67$ & -18\\
		\hline
		\end{tabular}
		\tablefoot{ 
		\tablefoottext{a}{Date of observation;}
		\tablefoottext{b}{GMVA elements with good data after calibration, imgaing and flagging: Effelsberg (Ef), Onsala (On), Mets\"ahovi (Me), Plateau de Bure (PdB), Very Long Baseline Array (VLBA, the eight stations equipped with 86\,GHz receivers) }
		\tablefoottext{b}{image noise level;}
		\tablefoottext{c}{peak flux density of the image;}
		\tablefoottext{d}{total flux density;}
		\tablefoottext{e}{major axis, minor axis and position angle of the restoring beam.}
		\tablefoottext{f}{The flux density might be underestimated by a about 22\% (see Sect. \ref{Data:GMVA})}
		}
		\end{table*}
		
		\begin{table}
			\centering
			\caption[]{Kinematic properties of the jet}
			\label{tab:Results:Kinematic}
			\begin{tabular}{cccc}
				\hline
				ID & $v_\mathrm{app,est}$ & $\beta_\mathrm{app,est}$ & $t_{0,\mathrm{est}}$ \\
				& [mas\,yr$^{-1}$] & [c] & [yr] \\
				\hline\hline
				C2  &  $-0.03\pm0.02$  &  $-0.09\pm0.06$  &    \\ 
				J1  &  $1.2\pm0.1$  	&  $4.0\pm0.3$  &  $2007.3\pm0.1$ \\ 
				J2  &  $1.3\pm0.1$  	&  $4.2\pm0.4$  &  $2007.5\pm0.1$  \\ 
				J3  &  $1.36\pm0.06$  	&  $4.5\pm0.2$  & $2007.67\pm0.04$  \\ 
				J4  &  $1.4\pm0.1$  	&  $4.5\pm0.4$  &  $2007.90\pm0.06$  \\ 
				J5  &	$1.4\pm0.1$ 	&  $4.5\pm0.5$	&  $2008.22\pm0.05$ \\ 
				\hline
			\end{tabular}
		\end{table}

\section{Discussion}
\label{Discussion}
	
	\subsection{Jet evolution and parameters}
	\label{Discussion:Evolution}
		
		The radio light curve at 230\,GHz (Fig.~\ref{fig:86Ghz-DistTimeLC}) reveals that the large outburst began in late 2006 with a small precursor flare occurring shortly before the main flux rise. 
		The top panel of Fig.~\ref{fig:86Ghz-DistTimeLC} shows the time evolution of the component distances with time and the back-extrapolated times when they emerged from the core.
		These ejection times coincide well with the characteristic evolution of the light curve at both frequencies. J1, J2 and J3 are the innermost jet components seen in our 2007 GMVA image (and also, at larger distances from the core, in later images). Their ejection seems related to the rise and peak of the light curve. The J1 ejection time falls close to a minor flare seen in the 230\,GHz light curve ($65\mathrm{\,d}\pm37\mathrm{\,d}$)\footnote{Based on the estimated ejection date and the flux density measurement at 230\,GHz on 2007 Feb 15}, while J2 and J3 are ejected close to the absolute maximum of the outburst ($-49\mathrm{\,d}\pm37\mathrm{\,d}$ and $-13\mathrm{\,d}\pm14\mathrm{\,d}$ for J2 and J3, respectively at 86\,GHz)\footnote{Based on the estimated ejection date and the flux measurement at 86\,GHz on 2007 Aug 02}. Here, the 230\,GHz light curve again reveals the presence of short-time-scale substructure. The ejections of J4 and J5 coincide with the decaying phase of the light curve in late 2007 and early 2008. While the 86\,GHz light curve does not show any specific event here, due to the limited sampling, the 230\,GHz data indicate local maxima which correspond well with the ejection times of these two components, in particular with J4.

		The apparent velocities are consistent with the fastest motions seen in previous observations at lower frequencies \citep{Jorstad2005,Kadler2008,Chatterjee2011,Lister2013,Beuchert2018,Jorstad2017}. These studies at 15\,GHz and 43\,GHz show a broader distribution of $\beta_\mathrm{app}$ between $\sim2c$ and $\sim8c$ while all our 86\,GHz components fall into a narrow range in apparent velocities between $(4.0 \pm 0.3) c$ and $(4.5 \pm 0.5) c$ (see Table~\ref{tab:Results:Kinematic}). This might be related to the short time range of our observations which might miss the full extent of activity in \threec. \cite{Chatterjee2011} analyzed the 43\,GHz observations shown here plus additional ones obtained over a period of 5.4 years, that include the time frame of the flare in 2007 and our GMVA observations. Our results are consistent with \cite{Chatterjee2011}, except for the association of their component K5 and the bend (J2, J3, in our model) which were not separated into individual components by \cite{Chatterjee2011}. 
		
		The critical angle of $\sim13\degr$ is slightly lower than the average value determined by \cite{Jorstad2017} of $16.3\degr\pm2.3\degr$ based on variability and consistent with the upper limit from \cite{Kadler2008} of $\leq21\degr$ based on 15\,GHz VLBA images. It was noted by \cite{Kadler2008} that these estimates of the angle are in contradiction with the lower limit of 21\degr determined for the kpc-scale jet by \cite{Lewis2005}. Intriguingly, \cite{Hogan2011} estimated the angle of the large scale jet to be $\sim 8.1\degr$ based on analysis of the X-ray jet in combination with the parsec-scale jet properties and assuming no deceleration. This is slightly lower than the previous estimates. However, the minimum Lorentz factor determined by \cite{Hogan2011} considering deceleration and bending fits our results, but is below the average value obtained by \cite{Jorstad2017} of $7.7\pm0.7$.

\begin{figure}
		\includegraphics[width=1\linewidth]{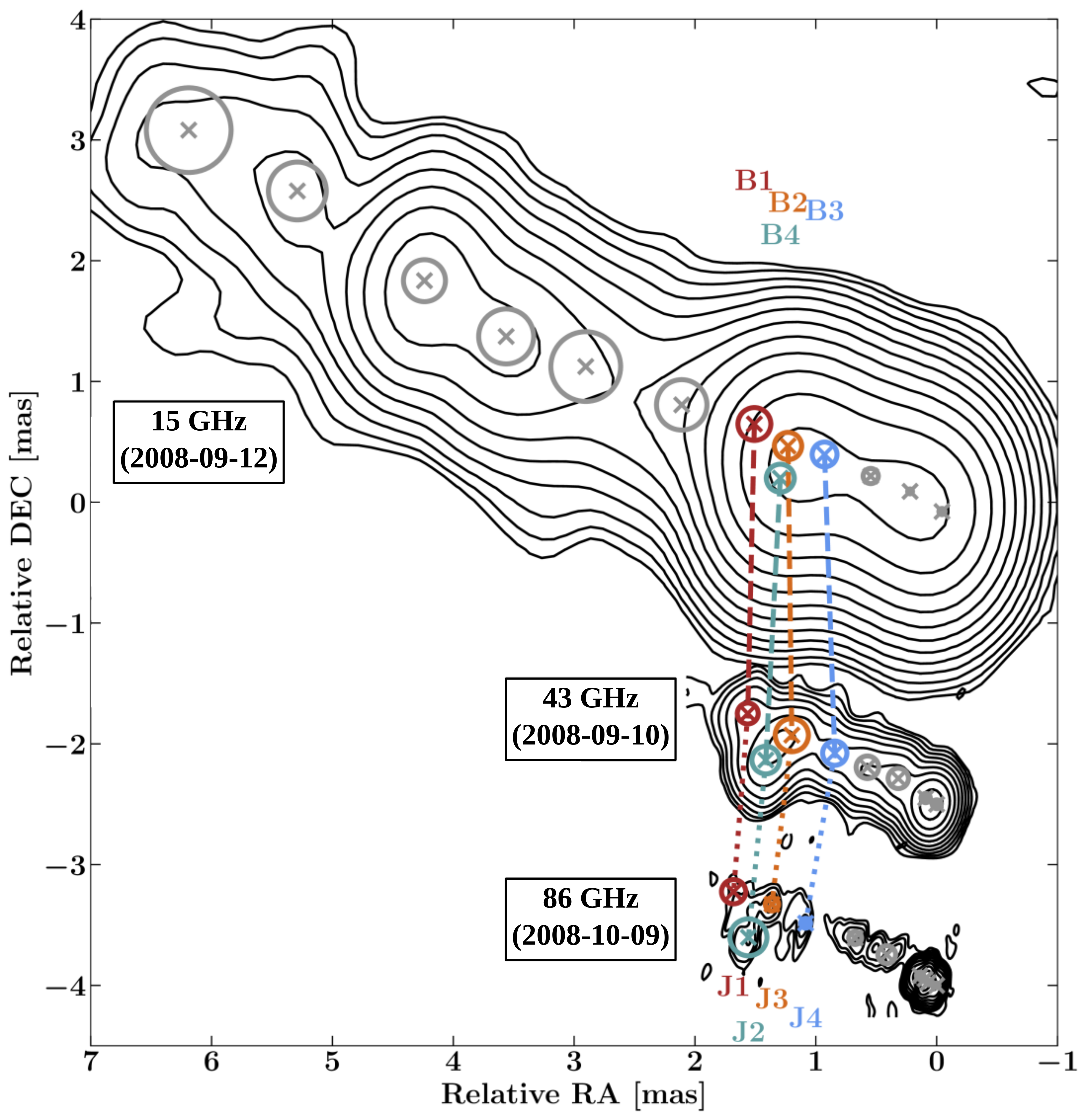}
		\caption{Multifrequency images of \threec at 15\,GHz (VLBA, from the MOJAVE programme), 43\,GHz (VLBA, from the Boston-Blazar-Group programme) and 86\,GHz (GMVA) to highlight the association of J1, J2, J3 and J4 with the 15\,GHz counterparts (B1, B2, B3, B4) reported in \cite{Beuchert2018}. In this image, we corrected for the core shift between 15\,GHz and 43\,GHz by applying a shift to the 15\,GHz image ($\Delta \mathrm{RA} = -0.22\mathrm{\,mas}$, $\Delta \mathrm{DEC} = -0.09$, \citealt{Beuchert2018}). The coordinates of the components are listed in Table \ref{tab:component_comparison}}
		\label{fig:15_43_86}
		\end{figure}

\begin{figure}
			\includegraphics[width=1\linewidth]{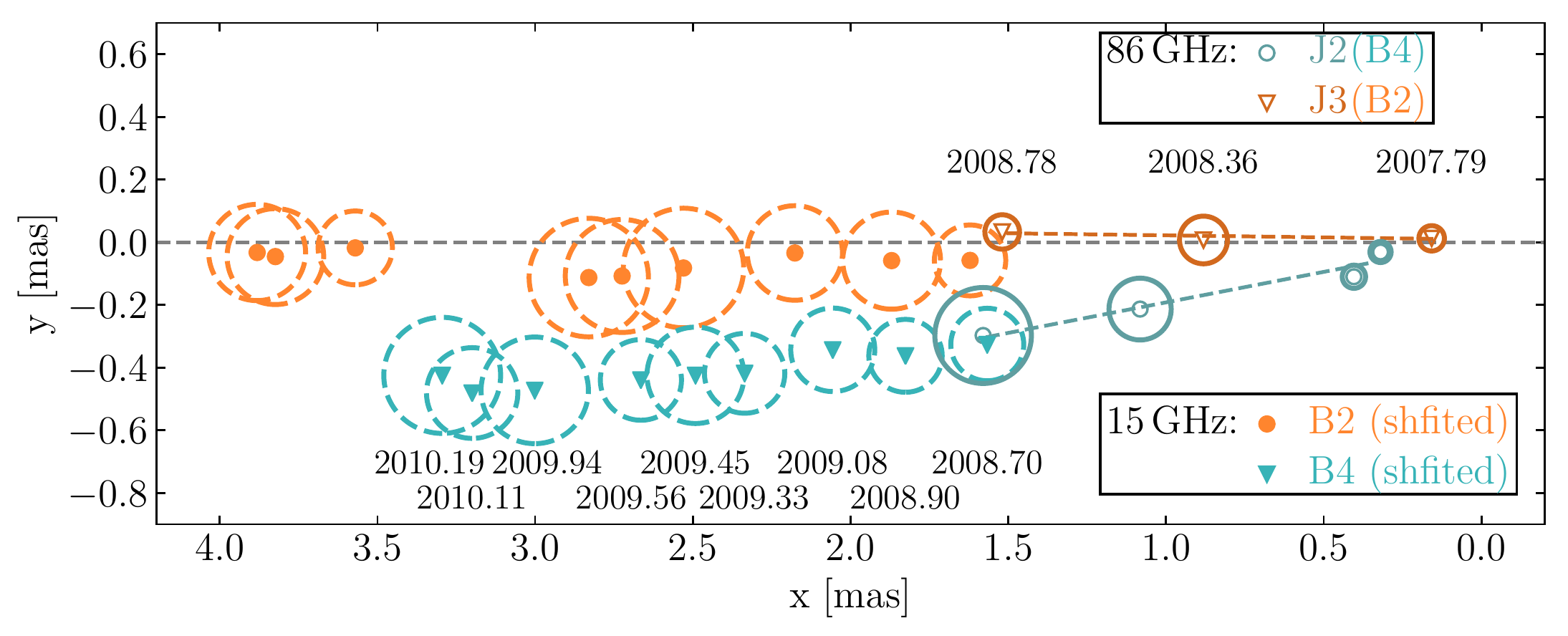}
			\caption{Modelfit components J2 and J3 at 86\,GHz and B2, and B4 at 15\,GHz from \cite{Beuchert2018} after rotation by the median position angle of the jet of 65\degr. B2 and B4 are corrected for the core shift between 15\,GHz and 43\,GHz. 
			}
			\label{fig:Rotated}
		\end{figure}

		Based on 15\,GHz VLBI data, \cite{Beuchert2018} report the appearance of a complex feature labeled `B' as the result of the 2007 outburst which is first detected about 1\,mas from the 15\,GHz core in their 15\,GHz VLBA analysis. It is shown to evolve into multiple components, B1, B2, B3 and B4, which seem to correspond to J1, J2, J3 and J4 at 86\,GHz as shown in Fig.~\ref{fig:15_43_86}.
		The 15\,GHz modelfit components are well aligned with the 43\,GHz components (see also Table \ref{tab:component_comparison}) with small differences being most likely the result of the different resolution. The difference in position between 43\,GHz and 86\,GHz components (see Table \ref{tab:component_comparison}) can be explained best by the movement of the jet features as show in Fig. \ref{fig:86Ghz-TimeEvo}.
		This consistency makes it possible to trace the future evolution of components J2 and J3 to larger distances than it would have been possible at 86\,GHz alone. This is illustrated in Fig.~\ref{fig:Rotated} (see also \citealt{Beuchert2018}). This plot shows the components J2 and J3 in combination with B2 and B4 from \cite{Beuchert2018} rotated by the median jet position angle of 65\degr and corrected for the core shift between 15\,GHz and 43\,GHz. 
		The core shift between 43\,GHz and 86\,GHz is very likely much smaller than the one from 15\,GHz to 43\,GHz as has been shown by other studies of AGN (e.g., \citealt{Lobanov1998a,Hada2011,Fromm2013b}). 

		Based on Fig.~\ref{fig:Rotated} it seems likely that J2/J3 at 86\,GHz and B2/B4 at 15\,GHz can be associated with each other. 
		This association has significant implication on the evolution of this jet structure which is discussed in Sect.~\ref{Discussion:Bend}.
		
	\subsection{Comparison with the outburst in 1996}

		As mentioned in Sect.~\ref{Intro} a major outburst was recorded in early 1996 similar to the one in 2007 in magnitude \citep{Alef1998}. 
		\cite{Alef1998} and \cite{Kharb2003} presented 43\,GHz VLBA images taken in 1996 July and 1996 Sep. The morphology indicates a small bend in the jet similar to the 43\,GHz images in 2008.04 shown in Fig. \ref{fig:86Ghz-TimeEvo}. In the 43\,GHz images obtained in 1998, \cite{Jorstad2005} detected a large diffuse structure located beyond $\sim2\mathrm{\,mas}$ from the core. The study concludes that this structure might represent the remnant of the 1996 outburst and measured $\beta_\mathrm{app}\approx6\mathrm{c}$ for its propagation. In addition, \cite{Jorstad2005} reported significant polarization in this feature at 43\,GHz. 

		The 43\,GHz images presented by \cite{Chatterjee2011} suggest that the feature related to the 2007 outburst has traveled a similar distance in a similar time range becoming increasingly diffuse and extended. At 15\,GHz, \cite{Beuchert2018} show that feature \textit{B} likewise becomes increasingly complex with time, eventually dominating the polarization of the entire jet. 
				
		These are all indications that the major outbursts in 1996 and 2007 represent similar events that differ from intermediate, minor flares from this radio source. Based on this, it seems probable that a similar event will occur in the future. If the \gray{} emission in \threec{} is correlated to the radio activity as has been seen in other sources, then a future large radio outburst would represent a prime opportunity to study the radio-\gray{} connection. \cite{Grandi2012} associated the \gray{} emission detected in \threec{} in 2008 Oct/Nov with a TS-value of 9.7
		with a small flare in late 2008 (see Fig. \ref{fig:86Ghz-DistTimeLC}). If this connection holds, then a larger radio outburst may coincide with a larger \gray{} event.
				
		Interestingly, components J4 and J5 which are trailing J1-J3 have a much lower flux density than the sum of bend-related components J1, J2 and J3 (Fig.~\ref{fig:86Ghz-FluxEvoComp}). This behavior is in agreement with the modeling of the 1996 flare in \cite{Kadler2008} and \cite{Perucho2008}. In those works, the authors suggested that the major flares are followed by drops in the mass flux being injected into the jet, resulting in a decrease in brightness. 
		Figure \ref{fig:86Ghz-FluxEvoComp} also indicates a steeper drop in flux density from the second to third epoch compared to the drop from the first to second epoch. This steeper drop cannot be explained by the flux density normalization of epoch 3 even if the systematic uncertainty is taken into account. However, because of the limited number of observations over the long time frame, we cannot undertake a detailed physical interpretation.
		
	\subsection{The nature of C1}
	\label{Discussion:Core}
	
		Traditionally, the brightest feature in the jet of \threec corresponded to the westernmost component. Here, we find an additional component upstream of the brightest emission region.
		Such a component was reported by \cite{Grandi2012} based on super-resolved 43\,GHz VLBA images. The study interpreted the brightest feature as the core of the jet and the component upstream of the core as the counter jet based on the stationary behavior, but no detailed analysis was performed.
			
\begin{figure}
		\includegraphics[width=1\linewidth]{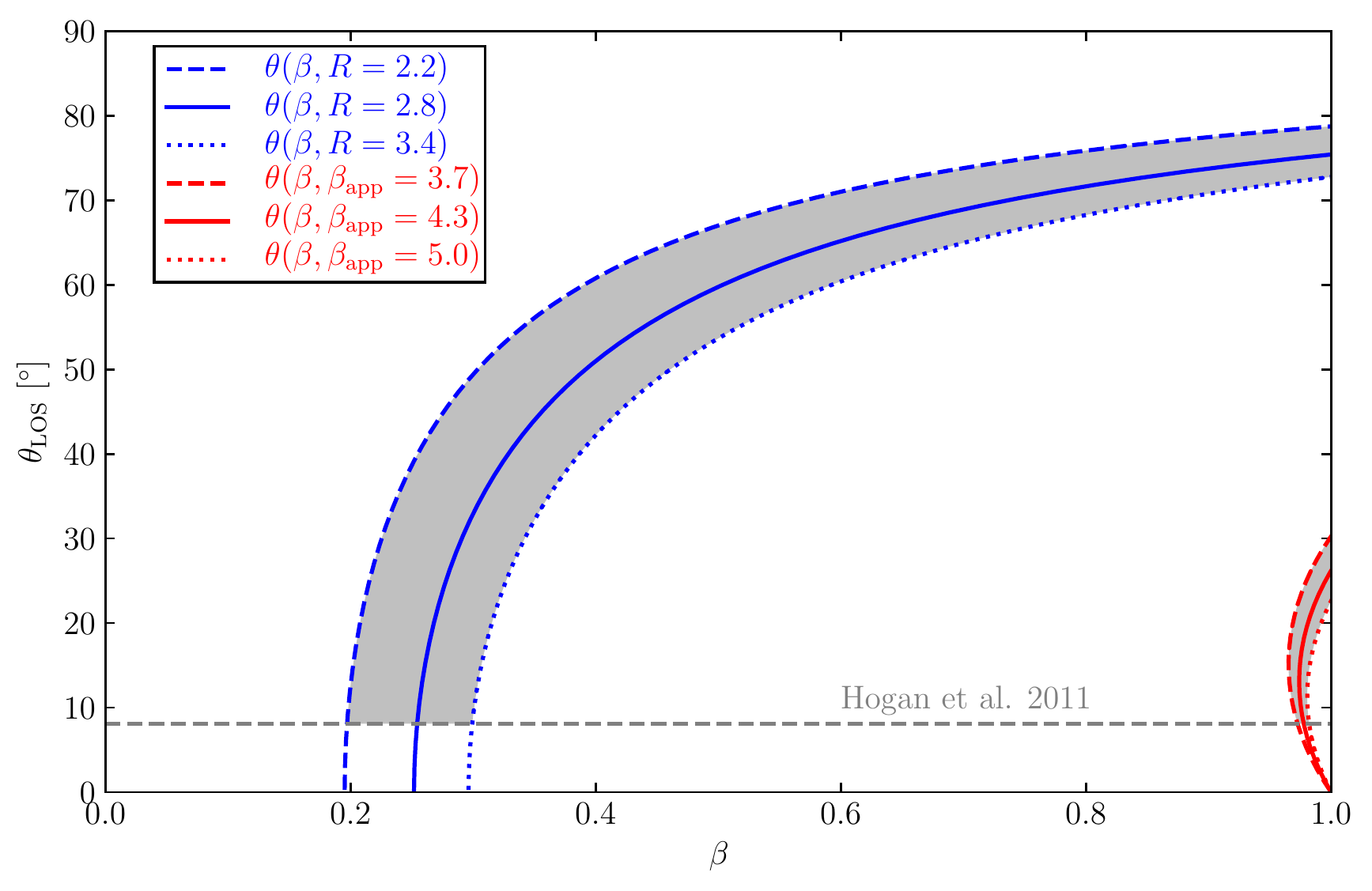}
		\caption{Angle between the jet and the line of sight $\theta_\mathrm{LOS}$ for a given intrinsic speed $\beta$, for the flux density ratio of the C1 and C2 components (blue) and the range of measured apparent velocities (red) of \threec. The gray shaded areas represent the parameter space for $\theta_\mathrm{LOS}$ and $\beta$ for both cases. The gray dashed line is a lower limit of the large-scale $\theta_\mathrm{LOS}$ from \cite{Hogan2011}.}
				\label{fig:86Ghz-Jet-angle}
		\end{figure}

		Various arguments seem to disfavor this scenario. In particular, using the ratio $R$ of the flux density of the jet and counter jet, it is possible to estimate $\theta_\mathrm{LOS}$ for a given intrinsic speed $\beta$ and a continuous, intrinsically symmetric jet (e.g., \citealt{Urry1995}) via
			\begin{equation}\label{eq:betatheta1}
			R = \left(\frac{1 + \beta\cos{\theta_\mathrm{LOS}}}{1 - \beta\cos{\theta_\mathrm{LOS}}}\right)^{2+\alpha}
			\end{equation}
		where $\alpha$ is the spectral index. Assuming C2 and C1 are actually the VLBI cores of the jet and counter jet, the corresponding flux densities in the first epoch yield $R\sim2.8$. The gray-shaded area marked by the blue lines in Fig. \ref{fig:86Ghz-Jet-angle} shows the resulting possible parameter space of $\theta_\mathrm{LOS}$ and $\beta$ accounting for the uncertainty of the flux-density calibration and the lower limit on $\theta_\mathrm{LOS}$ estimated by \cite{Hogan2011}. An independent way of constraining both parameters is using $\beta_\mathrm{app}$ from the previous section:
			\begin{equation}\label{eq:betatheta2}
			\beta_\mathrm{app} = \frac{\beta\sin\theta_\mathrm{LOS}}{1-\beta\cos\theta_\mathrm{LOS}} \quad .
			\end{equation}
		This yields the gray-shaded area between the red lines in Fig.~\ref{fig:86Ghz-Jet-angle} which shows no intersection with the gray-shaded area representing the allowed parameter space from Eq.~\ref{eq:betatheta1}. Solving this discrepancy would require a large change in $\theta_\mathrm{LOS}$ on scales below the resolution capabilities of the GMVA, which seems unlikely given that the parsec and kpc scale jet are well aligned and remarkably straight. 
		Alternatively, the scenario suggested by \cite{Grandi2012} would require unusually large intrinsic acceleration over the observed short distance.
		The reported angles and Doppler factors for \threec{} show that the radio emission is significantly boosted with $S_\mathrm{int} = \delta^{3+\alpha}S_\mathrm{obs}$, and any counter jet emission is expected to be de-boosted by  $1/\delta^{3+\alpha}$. For $\alpha=0$ this would require $\delta\sim 1.2$ to account for the given $R$ which is significantly below estimates from the kinematics of the jet.
		While we cannot entirely exclude the possibility of high intrinsic acceleration, we do not consider it to be the case here based on the arguments above. Therefore, we favor the interpretation that C2 represents a stationary component close to the core (C1) based on these arguments which is consistent with conclusions of \cite{Jorstad2017}
		This means that all modelfit components belong to the jet and no counter is visible which would correspond to much higher R value and a lower $\theta_\mathrm{LOS}$.

		In the canonical jet model \citep{Blandford1979,Koenigl1981}, the core corresponds to the transition region of the jet from synchrotron self-absorbed to optically thin emission where the opacity reaches $\tau=1$. Another possibility is that it represents a recollimation shock (e.g., \citealt{Daly1988,Gomez1995,Gomez1997,Fromm2011,Fromm2013b,Marscher2014}). In the case of components C1 and C2, a core-shift analysis is necessary in order to clarify whether C1 is a recollimation shock or not. In case of a recollimation shock, the core shift would not decrease with increasing frequency \citep{Fromm2018}. Future quasi-simultaneous multi-frequency VLBI observations up to 1\,mm wavelength might be able to test this.
			
		It is important to point out that our kinematic results only marginally depend on whether C1 or C2 was used for alignment as they are both kinematically stationary. Given the small distance between both components, we can use C1 as the reference point. 
		
	\subsection{The transversal structure of the jet}
	\label{Discussion:Bend}
	
	Figure~\ref{fig:Rotated} shows the distinct trajectory of component J2/B4 with respect to component J3/B2 and the link to other 15\,GHz components from \cite{Beuchert2018}. There is a clear connection between the components: 1) the 15\,GHz jet represents a smooth continuation of the 86\,GHz jet on larger scales, and 2) the long-term kinematics of the components are different, when the 15\,GHz evolution is considered, depending on the direction of propagation. Although components J2 and J3 show similar velocities at the observed 86\,GHz epochs, J2 decelerates with respect to J3 farther downstream. In relation with the originating flare, this process shows that the flare has substructure in space and time. 
			
	The growth of the observed jet bend might be interpreted as the development of a kink instability along the jet because of the flare \citep{Tchekhovskoy2016}. The structure is observed to grow out to a distance of $r\simeq3.5$\,mas at 15\,GHz, with a maximum amplitude of $0.5$\,mas. However, the 15\,GHz kinematics shows that this large amplitude does not trigger jet deceleration which would be expected following the disruption in the jet stream. In addition, the overall jet remains remarkably collimated on larger scales. Therefore, unless the instability and disruption develop at the boundary between the jet spine and the surrounding sheath, we regard the kink-instability option as improbable.  

	The 230\,GHz light-curve shown in Fig.\,\ref{fig:86Ghz-DistTimeLC} shows that the global evolution is modulated by short-scale spikes and drops. The five local maxima observed along the outburst can be related, within errors, to the ejection of the different modeled components. Components J1/B1 and J3/B2 propagate along the spine, whereas component J2/B4 moves along a  different PA (see Fig.~\ref{fig:Rotated}). Components J4 and J5 also propagate along the spine (at least out to a distance of 1.5\,mas covered by our GMVA data).
				
	Therefore, the ejection of bright components seems to happen in a discrete way when the launching region becomes active. 
	It is already known that the modeled radio components associated to bright regions propagating down the jet are ejected with different PAs (\citealt{Beuchert2018,Lister2013}) and that only stacked images combining a number of VLBI observations can reveal the whole time-averaged volume covered by the radio jet. Our observations probably reveal such behavior within a single flare. 
			
	The cross-identification of components from the 86\,GHz and the 15\,GHz images (see Fig. \ref{fig:Rotated}) allows us to follow the trajectory and kinematics of these components farther downstream of the 86\,GHz jet. We observe that component J2/B4 not only follows a trajectory along the bottom of the jet, but is also decelerated with respect to the components moving along the spine of the jet. It thus seems that J2/B4 is decelerating due to direct interaction with the ambient medium. 

	We can use the distance of J2 and J3 $r_\mathrm{J2,J3}$ in the last GMVA epoch as shown in Fig. \ref{fig:Rotated} including the size of the components to estimate the deprojected opening angle $\psi$ following $\tan\psi = r_\mathrm{J2,J3} / d_\mathrm{proj,J3} \times \sin\theta_\mathrm{LOS}$, where $d_\mathrm{proj, J3}$ corresponds to the projected distance of J3. With $8\degr \lesssim \theta_\mathrm{LOS}\lesssim 30\degr$, this yields $3.8\degr \lesssim \psi \lesssim 13\degr$, but the actual value may be larger due to possible further extended emission below the sensitivity limit. The range of $\psi$ is consistent with average values determined at 43\,GHz by \cite{Jorstad2017} using two different methods, i.e., $4.6\degr\pm 1.2\degr$ and $8.8\degr\pm6.2\degr$.

	In summary, the role that such peripheral components as J2/B4 play is two-fold. On the one hand, they carve the channel boundaries that allow the faster propagation along the jet spine. Additionally, they may be the origin of the jet transversal structure, i.e., of a spine-sheath structure within the radio-jet. We want to stress that such layers increase the jet stability (e.g., \citealt{Perucho2005,Perucho2007b,Marti2016,Perucho2019}).   

\section{Summary \& Conclusions}
\label{Summary}
	
	In this paper, we have presented three GMVA observations at 86\,GHz of the \gray loud radio galaxy \threec{} that were obtained over a period of one year shortly after a major outburst above 10\,Jy in 2007. The mm-wavelength light curves show that the whole flare covered a time range of more than a year starting in late 2006 and decaying well into 2008 with minor flares before and after the peak.
	
	The high dynamic range of the GMVA images reveal a complex morphology with a distinct feature in the form of a bend in the otherwise well collimated jet. The images and Gaussian components fitted to the data reveal an emission region upstream of the brightest feature of the jet. We consider the upstream emission as the core of the jet and not part of the counter jet. Both features are stationary and have high brightness temperatures of up to $3 \times 10^{11}\mathrm{\,K}$. One of them may be interpreted as part of a recollimation shock, but further multi-frequency VLBI observations at 86\,GHz and above are necessary. 
	
	We combine the GMVA data with 43\,GHz VLBA images to constrain the time evolution of the jet at 86\,GHz, which suggests that all features of the jet move with apparent velocities between $\sim4.0$\,c and $\sim4.5$\,c. The maximum speed corresponds to a critical angle of the jet to the line of sight of $\sim 13\degr$ and a critical Doppler factor of $\sim4.5$ consistent with previous studies. The high angular resolution of the GMVA observations enable us to separate the observed bend into two distinct features, J2 and J3, which have estimated ejection times that corresponds well to the maximum of the flare. Our results show that a kinematic study is possible even with the two observation windows per year of the GMVA for such a fast evolving jet as \threec if the GMVA observations are supported with interleaved 43\,GHz monitoring.

	We can associate the features of the bend down to 15\,GHz allowing us to track them over a larger distance. The VLBI data in combination with the radio light curves suggests that they originated from distinct ejections into the jet stream throughout the increased activity cycle of the AGN. The evolution of the bend indicates that we trace the transversal structure of the jet in terms of a spine and sheath as a result of the outburst. 
	
	To date, \threec has exhibited only one other major radio flare (in 1996) that matches the scale of this radio outburst in 2007. Only a much smaller radio outburst occurred in-between these two events.
	It is therefore plausible that a similar major event will take place in the future. Since a connection between a minor radio flare and the $\gray$ activity has been proposed in the literature \citep{Grandi2012}, it is interesting to consider that a future outburst similar to the one in 2007 could lead to increased high energy activity. Hence, \threec has the potential to be a key target source for studies that probe the origin and development of the smallest substructures in AGN jets that lead to the enigmatic short-term variability observed in some blazar jets. At an intermediate angle between classical blazars and radio galaxies and with its bright mm-band compact jet emission at small redshift, \threec offers the possibility for high-resolution multiwavelength studies of AGN-jet variability during high-energy flares.
	

\begin{acknowledgements}
We would like to thank the anonymous referee for helpful comments that improved the manuscript.
We would like to thank the internal MPIfR referee N. MacDonald for insightful comments that helped improve the manuscript. 
RS gratefully acknowledge support from the European Research Council under the European Union’s Seventh Framework Programme (FP/2007-2013)/ERC Advanced Grant RADIOLIFE-320745 and support by Deutsche Forschungsgemeinschaft grant WI 1860/10-1.
MP acknowledges financial support from the Spanish Ministry of Science through Grants PID2019-105510GB-C31, PID2019-107427GB-C33 and AYA2016-77237-C3-3-P, and from the Generalitat Valenciana through grant PROMETEU/2019/071.
I.A. acknowledges support by a Ram\'on y Cajal grant (RYC-2013-14511) of the "Ministerio de Ciencia e Innovaci\'on (MICINN)" of Spain. He also acknowledges financial support from MCINN through the “Center of Excellence Severo Ochoa” award for the Instituto de Astrof\'isica de Andaluc\'ia-CSIC (SEV-2017-0709). Acquisition and reduction of the POLAMI data was supported in part by MICINN through grant AYA2016-80889-P. IRAM is supported by INSU/CNRS (France), MPG (Germany) and IGN (Spain).
This research has made use of data obtained with the Global Millimeter VLBI Array (GMVA), which consists of telescopes operated by the MPIfR,
IRAM, Onsala, Metsahovi, Yebes, the Korean VLBI Network, the Green Bank Observatory and the Very Long Baseline Array (VLBA). The VLBA is a
facility of the National Science Foundation operated under cooperative agreement by Associated Universities, Inc. The data were correlated at
the correlator of the MPIfR in Bonn, Germany.
This study makes use of 43 GHz VLBA data from the Boston University gamma-ray blazar monitoring program (http://www.bu.edu/blazars/VLBAproject.html), funded by NASA through the Fermi Guest Investigator Program.
This research has made use of data from the MOJAVE database that is maintained by the MOJAVE team \citep{Lister2009b}.
The Very Long Baseline Array (VLBA) is an instrument of the National Radio Astronomy Observatory (NRAO). NRAO is a facility of the National Science Foundation, operated by Associated Universities Inc.
This research made use of the Interactive Spectral Interpretation System (ISIS) \citep{Houck2000} and a collection of ISIS scripts provided by the Dr. Karl Remeis observatory, Bamberg, Germany at http://www.sternwarte.uni-erlangen.de/isis/. This research made use of the NASA/IPAC Extragalactic Database (NED), which is operated by the Jet Propulsion Laboratory, California Institute of Technology, under contract with the National Aeronautics and Space Administration and of the VizieR catalogue access tool, CDS, Strasbourg, France.
\end{acknowledgements}

\bibliographystyle{aa}
\bibliography{References}

\begin{appendix}
	\section{Additional Tables}
	
	\begin{table*}
			\centering
			\caption[]{List of the GMVA stations, which participated in our observations in 2007 and 2008.}
			\label{tab:Data:GMVA}
			\begin{tabular}{cccccc}
				\hline
				Station & Diameter & SEFD & Epoch 1 & Epoch 2 & Epoch 3	\\
				&   [m]    & [Jy] & 2007-10-13 & 2008-05-11 & 2008-10-14 \\
				\hline
				\hline
				Mets\"{a}hovi & 14 & 17647 & \checkmark & \checkmark\tablefootmark{d} & \textsf{X}\\
				Onsala & 20 & 6122 & \checkmark & \checkmark & \checkmark\\
				Effelsberg & 100 & 929 & \checkmark & \checkmark & \checkmark \\
				Plateau de Bure & 35\tablefootmark{a} & 409 & \checkmark & \checkmark\tablefootmark{d} & \textsf{X}\\
				Pico Veleta & 30 & 643 & \checkmark\tablefootmark{c} & \checkmark & \checkmark\tablefootmark{e}\\
				8$\times$VLBA\tablefootmark{b} & 25 & 2941 & \checkmark & \checkmark & \checkmark\tablefootmark{f}\\
				\hline
			\end{tabular}
			\tablefoot{ 
				\tablefoottext{a}{Equivalent diameter of a phased interferometric array with 6$\times$15\,m telescopes.}
				\tablefoottext{b}{The eight VLBA stations are North Liberty, Fort Davis, Los Alamos, Pie Town, Kit Peak, Owens Valley, Brewster, Mauna Kea.}
				\tablefoottext{c}{No fringes after correlation.}
				\tablefoottext{d}{Flagged during a-priori calibration.}
				\tablefoottext{e}{Data were lost due to bad wheather.}
				\tablefoottext{f}{Data of North Liberty and Fort Davis were flagged during hybrid imaging.}}
	\end{table*}

		\begin{table}
			\centering
			\caption[]{Parameters of the GMVA jet components prior to positional alignment}
			\label{tab:Components}
			\begin{tabular}{c c c c c c }
				\hline
				ID\tablefootmark{a} & $S_\mathrm{\nu}$\tablefootmark{b} & $d$\tablefootmark{c} & $\phi$\tablefootmark{d} & $b_\mathrm{maj}$\tablefootmark{e} & $\log{T_\mathrm{b}}$\tablefootmark{f}\\
				& [Jy] & [mas] & [\degr] & [mas] & \\
				\hline\hline
				\multicolumn{6}{c}{2007-10-15 (2007.79)}\\
				\hline
				C1 & 1.51 & 0.070 & -128.6 & 0.044 & 11.14 \\
				C2a & 2.47 & 0.015 & -138.1 & 0.037 & 11.50 \\
				C2b & 1.80 & 0.039 & 42.9 & 0.046 & 11.18 \\
				J3 & 1.87 & 0.089 & 67.6 & 0.081 & 10.70 \\
				J2a & 1.77 & 0.256 & 75.8 & 0.067 & 10.84 \\
				J2b & 1.09 & 0.359 & 85.7 & 0.074 & 10.54 \\
				J1 & 1.53 & 0.590 & 76.8 & 0.120 & 10.27 \\
				\hline
				\multicolumn{6}{c}{2008-05-11 (2008.36)}\\
				\hline
				C1 & 0.47 & 0.104 & -117.9 & 0.049 & 10.54 \\
				C2 & 1.16 & 0.003 & -145.9 & 0.049 & 10.94 \\
				J5 & 0.55 & 0.089 & 59.4 & 0.089 & 10.08 \\
				& 0.33 & 0.357 & 59.9 & 0.150 & 9.41 \\
				J4 & 0.26 & 0.516 & 57.4 & 0.059 & 10.12 \\
				J3 & 0.88 & 0.778 & 64.9 & 0.152 & 9.82 \\
				J2 & 0.52 & 1.003 & 77.6 & 0.197 & 9.37 \\
				J1 & 0.14 & 1.297 & 67.4 & 0.250 & 8.61 \\
				& 0.06 & 4.108 & 65.9 & 0.166 & 8.60 \\
				\hline
				\multicolumn{6}{c}{2008-10-14 (2008.78)}\\
				\hline
				C1 & 0.66\tablefootmark{g} & 0.101 & -117.1 & 0.039 & 10.88 \\
				C2 & 0.74\tablefootmark{g} & 0.030 & -142.4 & 0.037 & 10.98 \\
				C2 & 1.44\tablefootmark{g} & 0.009 & 33.6 & 0.024 & 11.65 \\
				   & 0.38\tablefootmark{g} & 0.059 & 53.5 & 0.061 & 10.26 \\
				   & 0.11\tablefootmark{g} & 0.385 & 57.1 & 0.148 & 8.96 \\
				J5 & 0.10\tablefootmark{g} & 0.680 & 59.4 & 0.107 & 9.18 \\
				J4 & 0.06\tablefootmark{g} & 1.099 & 64.8 & 0.082 & 9.19 \\
				J3 & 0.10\tablefootmark{g} & 1.418 & 63.8 & 0.110 & 9.14 \\
				J2 & 0.17\tablefootmark{g} & 1.509 & 76.5 & 0.306 & 8.51 \\
				J1 & 0.05\tablefootmark{g} & 1.750 & 65.4 & 0.201 & 8.31 \\
				   & 0.06\tablefootmark{g} & 4.739 & 66.1 & 0.496 & 7.60 \\
				\hline
				\hline
			\end{tabular}
			\tablefoot{ 
				\tablefoottext{a}{ID of kinematic component}
				\tablefoottext{b}{Flux density of the component}
				\tablefoottext{c}{Distance of the component to center of the map}
				\tablefoottext{d}{Position angle of the component relative to the center of the map}
				\tablefoottext{e}{Major axis of the Gaussian component}
				\tablefoottext{f}{Logarithmic value of the brightness Temperatur}
				\tablefoottext{g}{As noted in Sect. \ref{Data:GMVA}, the flux density values obtained for the third VLBI measuremnt might be systematically underestimated by 22\%}
				}
		\end{table}

		\begin{table}
			\centering
			\caption[]{Coordinates of components at 15\,GHz (shift applied), 43\,GHz, and 86GHz used for Fig. \ref{fig:15_43_86}.}
			\label{tab:component_comparison}
			\begin{tabular}{cccc}
				\hline
				ID\tablefootmark{a} & $\nu$\tablefootmark{b} & Rel. RA\tablefootmark{c} & Rel. DEC\tablefootmark{d} \\
				& [GHz] & [mas] & [mas] \\
				\hline\hline
				B1 & 15 & 1.51 & 0.65 \\
				   & 43 & 1.56 & 0.75 \\
				J1 & 86 & 1.68 & 0.78 \\
				\hline
				B4 & 15 & 1.29 & 0.20 \\
				   & 43 & 1.41 & 0.37 \\
				J2 & 86 & 1.56 & 0.40 \\
				\hline
				B2 & 15 & 1.23 & 0.47 \\
				   & 43 & 1.20 & 0.57 \\
				J3 & 86 & 1.36 & 0.67 \\
				\hline
				B3 & 15 & 0.93 & 0.40 \\
				   & 43 & 0.84 & 0.42 \\
				J4 & 86 & 1.08 & 0.51 \\
				\hline
			\end{tabular}
			\tablefoot{ 
				\tablefoottext{a}{ID of kinematic component}
				\tablefoottext{b}{Frequency of the VLBI observation}
				\tablefoottext{c}{Relative Right Ascension of the component}
				\tablefoottext{d}{Relative Declination of the component}
				}
		\end{table}

		\begin{table}
			\centering
			\caption[]{86\,GHz flux density measurement from F-GAMMA and POLAMI}
			\label{tab:lightcurve_data}
			\begin{tabular}{cccc}
				\hline
				Date\tablefootmark{a} & Flux\tablefootmark{b} & Flux Error\tablefootmark{c} & Project\tablefootmark{d} \\
				$[$MJD] & [Jy] & [Jy] & \\
				\hline\hline
				54022.014 & 3.92 & 0.2 & POLAMI \\
				54113.504 & 4.11 & 0.07 & F-GAMMA \\
				54229.991 & 7.0 & 0.36 & POLAMI \\
				54236.013 & 7.38 & 0.38 & POLAMI \\
				54261.200 & 10.033 & 0.71 & F-GAMMA \\
				54312.617 & 12.444 & 0.258 & F-GAMMA \\
				54323.149 & 13.374 & 0.988 & F-GAMMA \\
				54332.994 & 13.46 & 0.7 & POLAMI \\
				54357.996 & 10.96 & 0.57 & POLAMI \\
				54359.018 & 12.83 & 0.66 & POLAMI \\
				54360.745 & 12.613 & 0.249 & F-GAMMA \\
				54382.479 & 12.057 & 0.224 & F-GAMMA \\
				54384.021 & 11.6 & 0.6 & POLAMI \\
				54447.249 & 10.482 & 0.51 & F-GAMMA \\
				54479.560 & 7.561 & 0.11 & F-GAMMA \\
				54507.508 & 6.898 & 0.199 & F-GAMMA \\
				54528.401 & 5.806 & 0.08 & F-GAMMA \\
				54588.072 & 4.055 & 0.158 & F-GAMMA \\
				54593.002 & 3.91 & 0.2 & POLAMI \\
				54616.812 & 3.903 & 0.066 & F-GAMMA \\
				54618.000 & 4.03 & 0.21 & POLAMI \\
				54627.992 & 4.08 & 0.21 & POLAMI \\
				54636.995 & 3.65 & 0.19 & POLAMI \\
				54642.977 & 3.373 & 0.206 & F-GAMMA \\
				54649.000 & 3.57 & 0.18 & POLAMI \\
				54675.730 & 3.619 & 0.137 & F-GAMMA \\
				54703.592 & 2.888 & 0.16 & F-GAMMA \\
				54720.004 & 3.21 & 0.17 & POLAMI \\
				54724.800 & 3.718 & 0.148 & F-GAMMA \\
				54746.537 & 4.061 & 0.215 & F-GAMMA \\
				54747.015 & 4.37 & 0.23 & POLAMI \\
				54777.355 & 6.71 & 0.757 & F-GAMMA \\
				54806.356 & 5.135 & 0.219 & F-GAMMA \\
				54830.024 & 4.02 & 0.21 & POLAMI \\
				54837.555 & 4.363 & 0.171 & F-GAMMA \\
				54882.005 & 3.84 & 0.2 & POLAMI \\
				54898.013 & 3.453 & 0.114 & F-GAMMA \\
				54963.848 & 2.171 & 0.109 & F-GAMMA \\
				55004.778 & 2.135 & 0.196 & F-GAMMA \\
			\end{tabular}
			\tablefoot{ 
				\tablefoottext{a}{Date of flux density measuremnt}
				\tablefoottext{b}{Flux density measurement}
				\tablefoottext{c}{Uncertainty of flux density measurement}
				\tablefoottext{d}{Flux density monitoring programme which obtained the measurement}
				}
		\end{table}

	\end{appendix}
\label{sec-appendix}

\end{document}